\documentclass[
    twocolumn,
	prd,
    amssymb,
	preprintnumbers,
	secnumarabic,
	nofootinbib,
	superscriptaddress]{revtex4-1}

\pdfoutput=1

\usepackage{graphicx}
\usepackage{enumitem}
\usepackage{latexsym}
\usepackage{amsfonts}
\usepackage{amssymb}
\usepackage{array}
\usepackage{pifont}
\usepackage{color}
\usepackage{xcolor}
\usepackage{amsmath}
\usepackage{slashed}
\usepackage{dcolumn}
\usepackage{verbatim}
\usepackage{float}
\usepackage{multirow}
\usepackage{xspace}
\usepackage[normalem]{ulem}
\usepackage{hyperref}
\usepackage{subfigure}
\usepackage{anyfontsize}
\usepackage{t1enc}

\definecolor{aquamarine}{rgb}{0.2,0.7,0.6}
\definecolor{cerulean}{RGB}{0,166,214} 
\definecolor{hypershade}{rgb}{0.3,0.3,0.8}
\definecolor{subtlered}{rgb}{0.8,0.3,0.3}

\hypersetup{
  pdfauthor={Nirmal Raj},
  pdftitle={one-hit wonder},
  pdfsubject={IISc Bag Model},
  colorlinks=true,
  citecolor=aquamarine,
  urlcolor=cerulean,
  linkcolor=gray
}

\newcommand{\gsim}{\gtrsim}
\newcommand{\lsim}{\lesssim}

\def\Oc{\mathcal{O}}

\newcommand{\beq}{\begin{eqnarray}}
\newcommand{\eeq}{\end{eqnarray}}
\newcommand{\bea}{\begin{eqnarray}}
\newcommand{\eea}{\end{eqnarray}}
\newcommand{\nn}{\nonumber}
\def\RNS{R_{\rm NS}}
\def\MNS{M_{\rm NS}}
\def\surftens{\sigma_{\rm ST}}
\def\rcrit{r_{\rm crit}}

\def\rhox{\rho_\chi}
\def\vx{v_\chi}
\def\mx{m_\chi}

\def\sigmaNx{\sigma_{\rm N\chi}}

\allowdisplaybreaks

\begin{document}

\title{Dark, deep, deconfining: \\ Phase transitions in neutron stars as powerful probes of hidden sectors}

\author{Aryaman Bhutani}
\email{aryamanb@iisc.ac.in}

\affiliation{Centre for High Energy Physics, Indian Institute of Science, C. V. Raman Avenue, Bengaluru 560012, India}

\author{Nirmal Raj}
\email{nraj@iisc.ac.in}

\affiliation{Centre for High Energy Physics, Indian Institute of Science, C. V. Raman Avenue, Bengaluru 560012, India}

\author{Zenia Zuraiq}
\email{zeniazuraiq@iisc.ac.in}

\affiliation{Department of Physics, Indian Institute of Science, C. V. Raman Avenue, Bengaluru 560012, India}

\date{\today}

\begin{abstract}

The interiors of neutron stars enjoy ideal conditions for the conversion of hadrons to a strange quark phase, theorized to be the stablest form of matter. 
Though numerous astrophysical means to prompt such a deconfinement phase transition have been suggested, they may be pre-empted by a large energy barrier for nucleation of quark matter droplets. 
We will show that interactions of hidden sectors of particles with nucleons may surmount the barrier if it exceeds deca-GeV energies, and spark a phase transition.
The neutron star would then, depending on the equation of state of QCD matter, convert to a black hole and/or set off a gamma-ray burst (GRB).
Using the observed existence of ancient neutron stars and estimates of the GRB rate, we then set some of the strictest (albeit conditional) limits on dark matter scatters, annihilations, and decays that are tens of orders stronger than those from terrestrial searches. 
For smaller energy barriers, lower limits on nucleon decay lifetimes of the order of $10^{64}$~yr may be obtained.

\end{abstract}

\maketitle

{\bf \em Introduction.}
 The quest for uncovering non-gravitational interactions of dark matter -- in the hope of revealing its enigmatic identity -- is in full force. 
 One distinctly promising approach of recent years has been to draw on the exceptional character of compact stars~\cite{BramanteRajCompactDark:2023djs,snowmass:ExtremeBaryakhtar:2022hbu}. 
 In this {\em Letter}, we will show that QCD deconfinement phase transitions in neutron stars could serve as an extraordinarily sensitive and unrivaled  
 probe of dark matter interactions with the visible sector.

As opposed to $^{56}$Fe~(with energy-per-baryon 930 MeV), 
strange matter -- chunks of $u$, $d$, $s$ quarks -- is believed to be the ground state of nuclear matter 
(with energy-per-baryon about 100 MeV smaller)~\cite{Bodmer:1971we,*Witten:1984rs,*FarhiJaffe:1984qu,*MadsenLecs:1998uh}.\footnote{See also Ref.~\cite{Holdom:2017gdc}, which obtains $ud$ matter as the ground state.}
An ideal environment for the QCD deconfinement phase transition (PT) from hadronic to strange matter to occur is the cold and dense interiors of neutron stars (NSs) -- corresponding to the low-temperature and baryon-rich regime of the QCD phase diagram.
In fact, such are the uncertainties of the equation of state (EoS) of high-density matter that it is unclear whether (all or some) NSs are actually ``strange stars'', a.k.a. ``quark stars'' (QSs) or ``hybrid stars''.
While there is some indirect evidence to suggest associating them with QSs~\cite{AnnalaQSEvidence:2019puf}, interpretations of measurements of their masses and radii by the currently operational NICER telescope favor neither hypothesis~\cite{NICERHybridStarsLi:2021sxb,*NICERTwinStars:Christian:2021uhd,NICERHuang:2025vfl}.

So, what induces the PT?
It is expected to be first-order~\cite{MapQCDPD:Rajagopal:1999cp,review:QCDPD:FukushimaHatsuda:2010bq}, implying the formation of droplets of quark matter in the NS core.
In turn, this implies overcoming, within a small pocket of hadron star material in the high-pressure region of the core, an energy barrier due to the surface tension of the droplet. 
Possible agents that could accomplish this may be~\cite{Alcock:1986hz,QBub:Olesen:1993ek,QBub:Harko:2004zz,review:QBubNucAstro:Bombaci:2016xuj}: 
the spin-down of the NS increasing 
the central temperature or pressure above the transition value, 
accretion of mass from a companion in a low-mass X-ray binary (LMXB),
binary NS mergers,
formation in high-pressure regions of either $ud$ matter that transitions to $uds$ matter or hyperons that cluster to seed strange matter, 
shock waves in the core-collapse supernova stage,
thermal nucleation of quark bubbles or burning of neutron matter in the hot proto-neutron star (PNS) medium,
lumps of strange matter formed elsewhere intersecting NSs and triggering conversions,
or neutrinos entering and interacting with the stellar core to inject energy. 
None of these mechanisms has been verified, and thus it's reasonable to suppose that at least some NSs are purely hadronic.
Our argument for this is that the transition energy barrier, which can range from $\Oc({10})$ to $\Oc({10^4})$~MeV, is simply too high for any of these mechanisms to effect a PT 
within the lifetime of the universe.
We then posit that dark matter could be a unique agent that can surmount this barrier, injecting the requisite energy via scatters, annihilations, or decays.
Depending on the high-density EoS, the PT could convert the NS to a black hole (BH) and/or produce a gamma ray burst (GRB). 
We will accordingly use the existence of ancient NSs and predicted GRB rates as criteria for setting limits on dark matter parameters.

NSs have been exploited as 
thermal detectors~\cite{NSvIR:Baryakhtar:DKHNS,NSvIR:Raj:DKHNSOps,NSvIR:Bell2019:Leptophilic,NSvIR:Riverside:LeptophilicShort,NSvIR:Riverside:Leptophiliclong,NSvIR:Bell:ImprovedLepton,NSvIR:GaraniHeeck:Muophilic,NSvIR:SelfIntDM,NSvIR:Hamaguchi:RotochemicalvDM2019,NSvIR:GaraniGenoliniHambye,NSvIR:Queiroz:Spectroscopy,NSvIR:Bell2020improved,NSvIR:DasguptaGuptaRay:LightMed,NSvIR:zeng2021PNGBDM,NSvIR:Queiroz:BosonDM,NSvIR:HamaguchiEWmultiplet:2022uiq,NsvIR:HamaguchiMug-2:2022wpz,NSvIR:SNeSBursts:Raj:2023azx,NSvIR:Hamaguchi:VortexCreepvDM2023,*NSvIR:ReheatedAll:Raj:2024kjq,NSvIR:NearestPulsars:Bramante:2024ikc,NSvIR:Bell2018:Inelastic,NSvIR:InelasticJoglekarYu:2023fjj,NSvIR:tidalfifthforce:Gresham2022,NSvIR:clumps2021,NSvIR:Bell:Improved,NSvIR:anzuiniBell2021improved,NSvIR:Marfatia:DarkBaryon,NSvIR:PseudoscaTRIUMF:2022eav,snowmass:ExtremeBaryakhtar:2022hbu,snowmass:Carney:2022gse} 
%%%%
or subjects of transmutations to BHs~\cite{Goldman:1989nd,Gould:1989gw,Bertone:2007ae,deLavallaz:2010wp,McDermott:2011jp,Kouvaris:2010jy,Kouvaris:2011fi,Kouvaris:2012dz,Bell:2013xk,Guver:2012ba,Bramante:2013hn,Bramante:2013nma,Kouvaris:2013kra,Bramante:2014zca,Garani:2018kkd,Kouvaris:2018wnh,Dasgupta:2020dik,Lin:2020zmm,Dasgupta:2020mqg,Fuller:2014rza,Bramante:2017ulk,Takhistov:2020vxs,Garani:2021gvc,Steigerwald:2022pjo,Bhattacharya:2023stq,Liang:2023nvo,Bramante:2024idl,snowmass:ExtremeBaryakhtar:2022hbu} to probe dark matter models.
The sensitivities so derived are extensive and greatly complement terrestrial searches.
We show that our method further improves on sensitivities by up to 60 orders of magnitude for dark matter masses above about 10~GeV.
We remark here that previous studies by and large make the same assumption about the composition of NSs as ours, {\em i.e.} that they are dominated by nucleons.
One further assumption we make is that nucleonic matter converts to quark matter through a first-order transition.

Dark matter-induced PTs in NSs has been studied before. Ref.~\cite{silkstrangeletsNS:2010xlt} obtained the minimum dark matter mass as a function of EoS parameters that would induce a deconfinement PT via self-annihilations of the dark matter captured by the NS. Ref.~\cite{DMann:bubblenucleation:Silk2019} further explores this scenario with variations of quark matter EoS and treatment of quantum tunneling through the energy barrier for the PT. Ref.~\cite{silkdecay:2014dra} induces the PT via decays (to various final states) of dark matter captured by the NS and identifies the decay lifetimes that would result in the PT. These studies discuss a GRB signal associated with the PT. Our work is a significant improvement over these in the following ways. While these references did not derive observational limits, we show bounds based on GRB rates and NS existence as mentioned above. On the dark matter side, we include scattering with nucleons as a PT trigger in addition to decays and self-annihilations; moreover, we obtain limits without assuming dark matter capture in NSs.

%%%%%%%
{\bf \em Deconfinement phase transitions.}
%%%%%%
We now discuss general requirements for setting up a hadron-to-quark PT in NSs (hereafter HS $\to$ QS) and proto-neutron stars (PNS $\to$ QS).

As in Ref.~\cite{review:QBubNucAstro:Bombaci:2016xuj}, we denote by $Q$* a temporary $ud$ phase formed from the transition that will then decay to the more stable $uds$ phase via weak processes.
The chemical potentials of hadronic ($\mu_{H}(P)$) and quark matter ($\mu_{Q^*} (P)$) as a function of pressure $P$ are given by their EoS models.
Now as we go deeper into the (P)NS, $P$ increases. 
Above a certain transition pressure $P_{\rm trans}$, $\Delta \mu \equiv \mu_{Q^*} - \mu_{H}$ becomes negative\footnote{It could flip sign again depending on the EoS, so that hadron matter may reappear deeper in the star~\cite{Ren:2020tll}.},
{\em i.e.} in a given volume, it is energetically more favorable to populate $Q^*$ matter than hadrons.
This does not guarantee a successful PT yet.
As it is first order, bubbles of $Q^*$ matter would need to nucleate at an energy cost of their surface tension $\surftens$.
These considerations may be expressed in terms of a potential barrier:
%%%%
\bea
\label{eq:potbarr}
U(r, P) &=& - \frac{4\pi r^3}{3} n_{Q^*} (P) |\Delta \mu (P)| + 4\pi r^2 \surftens~,\\
 \nn U_{\rm max} =&& 1675~{\rm MeV}~\bigg(\frac{\surftens}{100~{\rm MeV/fm}^2}\bigg)^3 \\
\nn && \times \bigg(\frac{{\rm fm}^{-3}}{n_{Q^*}}\bigg)^2  \bigg(\frac{100~{\rm MeV}}{|\Delta \mu|}\bigg)^2~, \\
\nn \rcrit =&&  3~{\rm fm}~\bigg(\frac{\surftens}{100~{\rm MeV/fm}^2}\bigg)
\bigg(\frac{{\rm fm}^{-3}}{n_{Q^*}}\bigg) \bigg(\frac{100~{\rm MeV}}{|\Delta \mu|}\bigg)~,
\eea
%%%%%
where $r$ is the size of a bubble fluctuation and $n_{Q^*}$ is the $Q^*$ number density.
(We have here neglected contributions from the curvature energy $\propto r$ and Coulomb energy barrier $\propto r^{-1}$~\cite{QBub:Harko:2004zz} that are generally ignored as being too small.)
The height of the energy barrier  $U_{\rm max}$ is obtained by setting $dU/dr = 0$ and the critical radius $\rcrit$ is the one above which $U < 0$, yielding a stable bubble.

While we have normalized the quantities in Eq.~\eqref{eq:potbarr} to somewhat typical values ({\em e.g.}, see Fig.~1 of Refs.~\cite{review:QBubNucAstro:Bombaci:2016xuj,DMann:bubblenucleation:Silk2019}), $U_{\rm max}$ is still highly uncertain.
The surface tension $\surftens$ is ill-understood since there is no single framework in QCD at high baryon density that models the potential of both nuclear and quark matter, so that only simple models are relied on. 
Estimates range from 10--150 MeV/fm$^2$~\cite{STestimate:Iida:1998pi,STestimate:Heiselberg:1992dx,STestimate:Lugones:2013ema,STestimate:Fraga:2018cvr}, with even 300~MeV/fm$^2$ given by dimensional analysis for an interface between nuclear matter and color-flavor-locked phase~\cite{STestimate:CFL:AlfordRajagopal:2001zr}.
The value of $\Delta \mu (P)$ depends on the poorly known 
EoS, as well as the point in the NS core at which the PT is triggered, and of $n_{Q^*}$ 
is expected to be $O({\rm fm}^{-3})$.
 For our main results we will use $U_{\rm max}~=~10$~GeV, the order of magnitude considered in Refs.~\cite{review:QBubNucAstro:Bombaci:2016xuj,silkstrangeletsNS:2010xlt,DMann:bubblenucleation:Silk2019}.
This is a conservative choice as only dark matter with mass $\mx \gsim U_{\rm max}$ is constrained as explained below. 
It also strengthens our null hypothesis that non-exotic energy injection mechanisms are out of play.
We will however discuss interesting scenarios with smaller $U_{\rm max}$ too.

Eq.~\eqref{eq:potbarr} implies that hadronic matter boils to make a stable $Q^*$ bubble if an energy $> U_{\rm max}$ were injected within a critical volume of radius $\rcrit$.\footnote{One can also in principle tunnel through the barrier for energy injections $< U_{\rm max}$, but this ``quantum nucleation'' timescale is exponentially higher and could be several orders greater than the age of the universe~\cite{review:QBubNucAstro:Bombaci:2016xuj}, as reflected in the dark matter parameter space probed in Ref.~\cite{DMann:bubblenucleation:Silk2019}. In this study we will restrict ourselves to energy injections $> U_{\rm max}$.} 
 The latter criterion is indeed met in our scenarios: the final states of our processes scatter inelastically with nucleons (with cross section, say, $\sigma_{\rm inel}$), initiating a shower of mostly pions.
From  $dE/dx = n_{\rm b} \sigma_{\rm inel} E$, the stopping length per factor of ten of energy degradation is 
$ X_{\rm had} = 0.2~{\rm fm}~({\rm fm}^{-3}/n_{\rm b}) (100~{\rm mb}/\sigma_{\rm inel})$. 
Hence, as QCD inelastic scatters are expected for $>$ GeV energies, the desired energy is deposited within $r_{\rm crit}$ for injections $>$ 10 GeV.\footnote{Note that Ref.~\cite{DMann:bubblenucleation:Silk2019} estimates stopping lengths using a formula from Ref.~\cite{Graham:2018efk} that applies only to ionic targets with bound nucleons.}
The resultant bubble will then grow because $U(r > \rcrit)$ is a decreasing function of $r$, so that energy is {\em released} with bubble growth, in turn triggering further conversion and setting up a chain reaction. 
 The conversion stops where the pressure drops below $P_{\rm trans}$. 
 As seen from Eq.~\eqref{eq:potbarr}, a smaller $U_{\rm max}$ could result in $r_{\rm crit} < X_{\rm had}$, in which case the bubble formation criterion is not met. However, Eq.~\eqref{eq:potbarr} implies that to get $r_{\rm crit} < X_{\rm had}$, the surface tension must be an order of magnitude smaller than the conventional value, or $n_{Q*}$ or $|\Delta\mu|$ should be much higher than typical values. Therefore, within the expected order of magnitude of the parameters in Eq.~\eqref{eq:potbarr} our energy injection criterion is satisfied.

{\bf \em Observational consequences.}
As the mass-radius configuration of the daughter star differs from that of the mother star, the difference in gravitational binding energy due to the conversion will be emitted in the form of neutrinos, gravitational waves (GWs) and electromagnetic radiation, in principle detectable for a(n unlikely) Galactic event~\cite{review:QBubNucAstro:Bombaci:2016xuj}.
A supernova-like shock and bounce is expected, resulting in a $10^{52}-10^{53}$~erg GRB-like event~\cite{review:QBubNucAstro:Bombaci:2016xuj}.
We can then place limits on dark matter by requiring that no NS has undergone a HS $\to$ QS GRB episode.
Since GRBs are thought to originate in core-collapse supernovae and NS binary mergers, our criterion is akin to demanding that the GRB rate due to our transition is smaller than the theoretically predicted one, $\sim 1/$galaxy/century~\cite{GRBRate:IzzardTout:2003uz}.
Note that this rate is of the order of the expected supernova rate. 

Another interesting possibility is that the PT converts the hadron star to a black hole.
Assuming reasonably that the {\em baryonic} mass $M_{\rm bary}$ is conserved during the conversion of the HS, a stable mass-radius configuration for the daughter/quark star would not result if the maximum {\em gravitational} mass $M_{\rm grav}$ is predicted by the EoS to be smaller than $M_{\rm bary}$~\cite{review:QBubNucAstro:Bombaci:2016xuj}.
In other words, the $M$-$R$ curve runs into a gravitationally unstable branch as can occur for a range of parameters in a given EoS that triggers a strong enough PT~\cite{Benic:2014jia,*LindblomWeber:2025wme}.
Such a ``failed quark star'' event (hereafter HS $\to$ BH) would particularly occur if 
the mother HS is on the heavier side~\cite{review:QBubNucAstro:Bombaci:2016xuj,NICERHybridStarsLi:2021sxb,*NICERTwinStars:Christian:2021uhd}.
The existence of heavy NSs could thus be used to set bounds on PT-triggering dark matter.
We note here that this class of conversions too would result in GRB + neutrino + GW signatures, as presumably a ``proto-quark star'' configuration is first formed that then collapses to a BH.
In this study we will display results demanding a concrete criterion: no HS $\to$ BH energy deposition event by dark matter over the lifetime of PSR J0952$-$0607 (taken here as its spin-down age of $\tau_{\rm NS}$ = 4.9 Gyr~\cite{NSHeaviestAge:Nieder:2019cyc}).
It is the heaviest known NS with mass~\cite{NSHeaviest:Romani:2022jhd} and radius~\cite{NSHeaviestPossibleRadius:2025edc}
%%%%
\bea
\MNS &=& 2.35~M_\odot~, \ \ \ \ \RNS = 12~{\rm km} \\
\nn &\Rightarrow& \ v_{\rm esc} =\sqrt{\frac{2 G\MNS}{\RNS}} = 0.76~c~.
\label{eq:benchmarkNS}
\eea
%%%%
Other known heavy, $\Oc$(Gyr)-old NSs~\cite{NSHeavy2Pt08:Fonseca:2021wxt,*NSHeavy1Pt97:Demorest:2010bx} may also be used, yielding near-identical results.
Note that the benchmark NS is here required to be a HS, with other NSs allowed to be QSs.
Note that the benchmark NS is here required to be a HS, with other NSs allowed to
be QSs. The assumption that heavy NSs are hadronic is realized in multiple EoS models~\cite{review:Reddy:2002ri,review:QBubNucAstro:Bombaci:2016xuj,NICERTwinStars:Christian:2021uhd}.
Interestingly, this criterion is roughly equivalent to the GRB rate criterion discussed above.
Since the distribution of NS ages $t_{\rm NS}$ is weighted as $t_{\rm NS}/t^{\rm max}_{\rm NS}$, where $t^{\rm max}_{\rm NS} = \Oc(1-10)$~Gyr is the age of the oldest NSs, our GRB rate per NS must simply be $\lesssim 1/t^{\rm max}_{\rm NS}$: approximately the HS $\to$ BH criterion.
A more sophisticated estimate of the GRB rate may be made accounting for dark matter and NS density and velocity variations across galaxies, NS mass-radius distributions, and so on, but we expect the equivalence to hold since the HS $\to$ BH criterion applies to a region with a typical dark matter density in a typical galaxy.
Note that the GRB rate criterion requires all NSs to be HSs.
Thus our two criteria make two extreme assumptions about the content of NSs, yet due to the equivalence above, interpolating between these assumptions would result in similar limits.

We can now define a ``deconfinement radius'' $R_{\rm dcf}$ in the NS core within which $P > P_{\rm trans}$.
As $P_{\rm trans}$ is usually some typical pressure in the NS~\cite{review:QBubNucAstro:Bombaci:2016xuj}, $R_{\rm dcf}$ is expected to be an $\Oc(1)$ fraction of $\RNS$ as depicted in, {\em e.g.} Refs.~\cite{PTSims:Herzog:2011sn,AnnalaQSEvidence:2019puf}.
For showing our results, we will take $R_{\rm dcf} = \RNS/2 = 6$~km.

The two criteria above assume that at least some NSs are hadron stars. 
This premise can be relaxed if we consider conversions of the post-supernova PNS to a BH~\cite{PNSPT:Pons:2001ar,PNSPT:Bombaci:2009jt,PNSPT:Bombaci:2011mx, review:QBubNucAstro:Bombaci:2016xuj}.
As PNSs last only $\Oc(10-100)$~s, this will result in bounds that are weaker than from HS conversions, nonetheless are more robust as the matter in a PNS is certainly expected to be initially hadronic.
Incidentally, a PNS $\to$ BH event in the next Galactic supernova can be picked up in neutrino signals~\cite{PNSLateTimenus:LiBeacom:2020ujl}.
We will re-use NS potential barrier parameters for the PNS as finite-temperature EoS effects ({\em e.g.} Ref.~\cite{PNSEoS:ReddyPrakashLattimer:1998mm}) yield $U_{\rm max}$ of the same order as Eq.~\eqref{eq:potbarr}~\cite{review:QBubNucAstro:Bombaci:2016xuj}.
For concreteness, we will take the deconfinement timescale $\tau_{\rm PNS} = 10$~s and the PNS deconfinement radius as $R_{\rm PNS}/2 = 5\RNS/2 = 30$~km.

%%%%%
\begin{figure}[!h]
    \centering
    \includegraphics[width=0.99\linewidth]{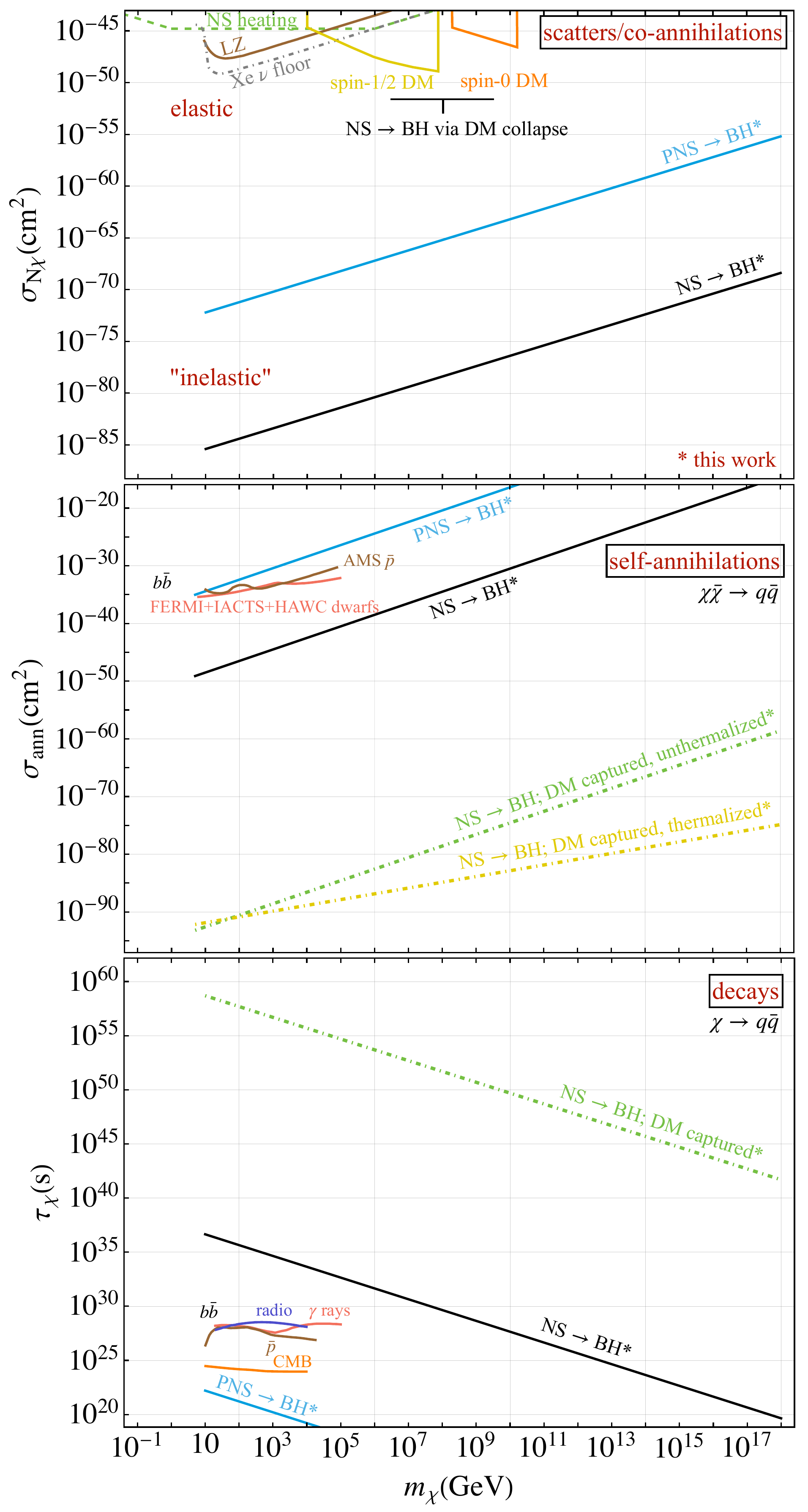}
    \caption{
Limits on our dark matter interaction scenarios, assuming a hadron $\to$ quark first-order phase transition occurs in neutron stars, with $U_{\rm max} = 10$~GeV. 
The scattering limits apply to dark matter models that prompt deep inelastic scatters, co-annihilations, and self-destruction, though we also show for comparison constraints on elastic scatters from a direct search at LZ~\cite{LZ:2024zvo},
the neutrino floor at xenon-based direct detectors where search sensitivities become limited~\cite{Billard:2013qya},
limits on non-annihilating dark matter that captures in neutron stars and transmutes them to black holes, and the reach of future observations of neutron stars overheated by dark matter~\cite{BramanteRajCompactDark:2023djs}. 
Indirect detection limits are taken from Ref.~\cite{reviewDMCirelliStrumiaZupan:2024ssz}; we have divided the excluded $\langle \sigma v\rangle$ values by the typical halo speed $10{^{-3}}c$; these can in principle be extended to higher masses by suitable recasts~\cite{snowmass:Carney:2022gse}.  
In principle there may be limits below $\mathcal{O}$(10) GeV mass if $U_{\rm max}$ were smaller.}  
    \label{fig:limits}
\end{figure}
%%%%

%%%%%
{\bf \em Dark matter as the agent.}
For a halo dark matter density $\rho_\chi$ and typical speed $v_\chi$, the mass flow rate of dark matter particles through our NS is~\cite{Goldman:1989nd}
%%%%%
\bea
\label{eq:massflowrateNS}
\dot{M}_{\rm \chi flow} &=&  \rhox \vx \times \pi  b_{\rm max}^2 = 2 \times 10^{25}~{\rm GeV}/{\rm s}~,\\
\nn  b_{\rm max} &=& \gamma \RNS (v_{\rm esc}/\vx)~,
\eea
%%%%%
where the second equality is for the solar neighborhood values $\rhox = 0.3$~GeV/cm$^3$ and $\vx = 10^{-3}$~c, and the factor $\gamma = (1-v^2_{\rm esc})^{-1/2} = 1.53$ accounts for the stellar radius as apparent to distant observers.
The number flow rate is
$\dot{N}_{\rm \chi flow} = \dot{M}_{\rm \chi flow}/\mx$.
The gravitational focusing factor in Eq.~\eqref{eq:massflowrateNS} enhances not only the dark matter flux but also the dark matter density within the NS with respect to the halo value, obtained from the continuity equation as
%%%%
\beq
\label{eq:DMdensityenhanceNS}
\rho_{\chi}^{\rm NS} = \rhox \bigg( \frac{v_{\rm esc}}{\vx} \bigg)~,
\eeq
%%%%
neglecting the slight variation in escape speeds in the NS interior.

For the flux through the deconfinement volume we replace $\RNS \to R_{\rm dcf}$ above.
The rate at which incident dark matter particles scatter or co-annihilate with NS nucleons in this volume in the optically thin limit is
%%%%%
\bea
\label{eq:ratescat}
 \Gamma^{\rm NS}_{\rm scat} &=& \dot{N}_{\rm \chi flow} \times \sigmaNx/\sigma_{\rm geo}~ \\
\nn &\simeq& \frac{5}{\rm Gyr} \bigg(\frac{\sigmaNx}{10^{-84}~{\rm cm}^2} \bigg) \bigg(\frac{\rm 10~GeV}{\mx} \bigg)
\bigg(\frac{R_{\rm dcf}}{6~{\rm km}}\bigg)^3~,
\eea 
%%%%%
with the geometric cross section $\sigma_{\rm geo}~=~3 (m_n/\rho_{\rm NS})/(4 R_{\rm dcf}) \simeq 3 \times 10^{-45}~$cm$^2$.

For $U_{\rm max}$ = 10 GeV, elastic scatters may not transfer enough energy to nucleon targets to trigger a PT~\cite{NSvIR:Baryakhtar:DKHNS}.
However, deep inelastic scattering with rates not too smaller than elastic scatters are expected for many dark matter interaction structures, particularly for NSs with high escape speeds such as ours~\cite{NSvIR:anzuiniBell2021improved}.
For $\mx > 10$~GeV there is a finite cross section for the entire mass energy to be deposited by breaking nucleons.
States with negative baryon number, such as ``dark anti-neutrons''~\cite{NSvIR:Marfatia:DarkBaryon,darkantin:Keung:2019wpw} and those arising in ``hylogenesis''~\cite{NSvIR:hylogenenesis1:2010,*NSvIR:hylogenenesis2:2011} could co-annihilate with nucleons and release mass energy.
``Self-destructing'' dark matter could deposit its mass energy upon scattering~\cite{selfdestructDM:Grossman:2017qzw}.
We take $\sigmaNx$ in Eq.~\eqref{eq:ratescat} to denote the net cross section that deposits energy $> U_{\rm max}$ in any one of these scenarios.

Dark matter particles passing through the NS may also undergo self-annihilations or decays to Standard Model final states that deposit energy above $U_{\rm max}$.
For simplicity we assume quark-antiquark final states to initiate an inelastic QCD cascade; other final states involving (anti)leptons and bosons would produce electronuclear and photonuclear showers with greater stopping lengths than $X_{\rm had}$~\cite{Graham:2018efk}, a detailed study of which we leave to future work. 
For cross section $\sigma_{\rm ann}$, the rate of annihilation is
%%%%
\bea
\label{eq:annratenocap}
\Gamma_{\rm ann}^{\rm NS}  &=&  \left(\frac{\rho_{\chi}^{\rm NS}}{\mx} \right)^2  \sigma_{\rm ann} v_{\rm esc} \left(\frac{4\pi}{3} R_{\rm dcf}^3 \right)~\\
\nn &=& \frac{0.06}{\rm Gyr}  \bigg( \frac{\sigma_{\rm ann}}{10^{-50}~{\rm cm}^2} \bigg) \bigg( \frac{10~{\rm GeV}}{\mx} \bigg)^2 \bigg( \frac{R_{\rm dcf}}{6~{\rm km}} \bigg)^3~,
\eea
%%%%
and for a dark matter lifetime of $\tau_\chi$, the rate of decay is
%%%%%%
\bea
\label{eq:decayratenocap}
\Gamma_{\rm decay}^{\rm NS} &=&  \left(\frac{\rho_\chi^{\rm NS}}{\mx} \right) \left(\frac{4\pi}{3} R_{\rm dcf}^3 \right) \tau^{-1}_\chi~\\
\nn &=&  \frac{1}{{\rm Gyr}}  \bigg( \frac{10^{36}~{\rm s}}{\tau_\chi} \bigg) \bigg(\frac{10~{\rm GeV}}{\mx} \bigg) \bigg( \frac{R_{\rm dcf}}{6~{\rm km}} \bigg)^3~,
\eea
%%%%%%
where the second lines in Eqs.~\eqref{eq:annratenocap} and \eqref{eq:decayratenocap} assume our benchmark NS and local dark matter density.

Now, suppose the flux of dark matter passing through the NS (Eq.~\eqref{eq:massflowrateNS}) is captured in its gravitational potential, as can happen via scattering on the stellar constituents.
Should such capture occur via elastic scatters, or via dark matter scattering inelastically to an excited state, a PT is not triggered as mentioned.
 If the rate of dark matter self-annihilation or decay is negligible enough, stellar capture can then steadily increase the population of dark matter inside the NS as 
%%%%
\beq
N_\chi^{\rm NS} (t) = \dot{N}_{\rm \chi flow} t~
\label{eq:NSDMpopvstime}
\eeq
%%%%
assuming all incident dark matter captures.
Within a volume $V$ in the NS, the self-annihilation rate is now 
%%%%
\bea
\Gamma_{\rm ann}^{\rm NS, cap} (t) &=&  \bigg( \frac{N_\chi^{\rm NS}(t)}{V} \bigg)^2 \times V \times \sigma_{\rm ann} v_\chi^{\rm NS}~\\
\nn &=& \frac{0.8}{{\rm Gyr}}  \bigg( \frac{\sigma_{\rm ann}}{10^{-92}~{\rm cm}^2} \bigg) \bigg( \frac{10~{\rm GeV}}{\mx} \bigg)^2  \\
\nn && \bigg( \frac{6~{\rm km}}{R_{\rm dcf}} \bigg)^3 \bigg( \frac{v_\chi^{\rm NS}}{v_{\rm esc}}\bigg) \bigg(\frac{t}{\tau_{\rm NS}} \bigg)^2~,
\label{eq:annrateyescap}
\eea
%%%%
where $v_\chi^{\rm NS}$ is the dark matter speed in the NS, and the decay rate is 
%%%%%%
\bea
\label{eq:decayrateyescap}
 \Gamma_{\rm decay}^{\rm NS, cap} (t) &=& \bigg( \frac{N_\chi^{\rm NS}(t)}{V} \bigg) \times V \times \tau^{-1}_\chi~\\
\nn &=& \frac{1}{{\rm Gyr}}~ \bigg( \frac{10^{58}~{\rm s}}{\tau_\chi}\bigg)  \bigg(\frac{10~{\rm GeV}}{\mx} \bigg) \bigg( \frac{t}{\tau_{\rm NS}}\bigg) ~. 
\eea
%%%%%%

For self-annihilations, the choices of  $v_\chi^{\rm NS}$ and $V$ depend on whether dark matter has, via repeated scattering, thermalized with the NS material post-capture, whereas for decays, as Eq.~\eqref{eq:decayrateyescap}
is independent of them, thermalization is irrelevant. 
The timescale for thermalization is highly model-dependent~\cite{NSvIR:ThermaliznBertoni:2013bsa,NSvIR:GaraniGuptaRaj:Thermalizn,*NSvIR:BellThermalize:2023ysh}. 
If it is shorter than the self-annihilation time, we can take $v_\chi^{\rm NS} = \sqrt{3 T_{\rm NS}/\mx}$ and $V = 4\pi r_{\rm th}^3/3$, the ``thermal volume'' within which one expects most of the dark matter to have settled, where~\cite{BramanteRajCompactDark:2023djs}
%%%%
\beq
r_{\rm th} \simeq 20~{\rm cm}~\bigg( \frac{{\rm GeV}}{\mx} \bigg)^{1/2}  \bigg( \frac{T_{\rm NS}}{10^3~{\rm K}} \bigg)^{1/2}~.
\eeq
%%%%
For dark matter unthermalized at the time of self-annihilation, we take $v_\chi^{\rm NS} \simeq v_{\rm esc}$ and $V = 4\pi R_{\rm dcf}^3/3$.
Partially thermalized dark matter, not considered here, would interpolate between these scenarios.
Thanks to the tiny $\sigma_{\rm ann}$ in the problem, for some parameters thermalized dark matter may collapse to form a star-destroying black hole~\cite{BramanteRajCompactDark:2023djs}, but this may be averted in several models that yield thermalization timescales exceeding the NS lifetime~\cite{NSvIR:GaraniGuptaRaj:Thermalizn,*NSvIR:BellThermalize:2023ysh}.

{\bf \em Results.}
Fig.~\ref{fig:limits} shows our limits, obtained by the HS $\to$ BH criterion for the NS in Eq.~\eqref{eq:benchmarkNS} (corresponding also to order-of-magnitude limits from the GRB rate, as discussed) and the PNS $\to$ BH criterion. 
We stress that the validity of these limits is contingent on the assumptions that hadron matter populates NSs and that it undergoes first-order PT to quark matter. 
While these are plausible by consensus, direct evidence for them is lacking at present, and thus our limits hinge on the QCD community bearing out our assumptions.

The inelastic scattering/co-annihilation limits on the dark anti-neutron are obtained by demanding $\Gamma_{\rm scat}^{\rm NS} \tau_{\rm NS} < 1$ with the kinematic requirement of $\mx > U_{\rm max}$, as a single interaction would deposit PT-triggering energy.
Also shown are limits and sensitivities from NS heating, conversions of NSs to BHs via dark matter capture and collapse (applicable only to non-annihilating dark matter models), and direct terrestrial searches.
We probe cross sections tens of orders of magnitude smaller than the other setups due to the enormous integrated flux of dark matter through our system, out of which only one scatter is required.
Though these other setups assume elastic scatters, we expect their bounds to be of roughly the same order of magnitude, even for the inelastic scatters we consider, as these are set mainly by their exposures.

The self-annihilation HS $\to$ BH limits are obtained from the conditions $\Gamma^{\rm NS}_{\rm ann} \tau_{\rm NS} < 1$ and $\Gamma^{\rm NS, cap}_{\rm ann} \tau_{\rm NS} < 1$ and PNS $\to$ BH limits from $\Gamma^{\rm NS}_{\rm ann} \tau_{\rm PNS} < 1$, with the condition $2 \mx > U_{\rm max}$. 
Similarly the decay limits are obtained from the conditions $\Gamma^{\rm NS}_{\rm decay} \tau_{\rm NS} < 1$ and $\Gamma^{\rm NS, cap}_{\rm decay} \tau_{\rm NS} < 1$, and $\Gamma^{\rm PNS}_{\rm decay} \tau_{\rm PNS} < 1$, with $\mx > U_{\rm max}$.
To be safe from direct search limits, we assume that NS capture at geometric cross sections has occurred via inelastic scattering to an excited state~\cite{NSvIR:Baryakhtar:DKHNS,NSvIR:Bell2018:Inelastic,NSvIR:InelasticJoglekarYu:2023fjj}, or velocity-dependent scattering~\cite{NSvIR:Raj:DKHNSOps,NSvIR:Bell:Improved,NSvIR:PseudoscaTRIUMF:2022eav}, that is inaccessible to terrestrial detectors. 
The annihilation limits for captured-but-unthermalized dark matter are 44 orders stronger than the uncaptured case because the dark matter population in the same annihilation volume is $10^{22}$ times greater in the first scenario while the annihilation rate goes as the square of the number density (Eqs.~\eqref{eq:DMdensityenhanceNS} and \eqref{eq:NSDMpopvstime}).
From Eq.~\eqref{eq:annrateyescap} and the discussion below it, the ratio between the annihilation rates for the unthermalized and thermalized scenarios for a given $\sigma_{\rm ann}$ is  
%%%%
\beq
\nn \bigg( \frac{r_{\rm th}}{R_{\rm dcf}} \bigg)^3 \bigg(v_{\rm esc} \sqrt{\frac{\mx}{3 T_{\rm NS}}} \bigg)~ \propto \frac{T_{\rm NS}}{\mx}~,
\eeq
%%%%
resulting in the difference in slopes in the limits in Fig.~\ref{fig:limits}. 
In particular, the limits are stronger at higher $\mx$ for the thermalized scenario.
As discussed, the decay limits do not depend on the thermalization volume or dark matter speed, and hence the thermalized and unthermalized scenarios are scaled with respect to the uncaptured scenario by a factor of the dark matter population in the NS, $\sim 10^{22}$.
These limits near $\mx$~=~10~GeV are within a few orders to those obtained in Ref.~\cite{silkdecay:2014dra}, though the limits on $\tau_\chi$ in that work appear to be erroneously displayed as $\mx$-independent.
Our self-annihilation and decay limits are several orders of magnitude stronger than Earth-based indirect searches due to the cosmological-scale exposure of our compact star detector confronting a single event as opposed to the quite finite exposures of instruments collecting a large flux of particles.

{\bf \em Discussion.}
In this work, we used neutron stars' non-conversion to black holes and non-production of excess GRBs to set conditional limits on dark matter that could trigger a first-order deconfinement phase transition in their cores.
Whilst we derived stringent bounds on dark matter interactions, our treatment may also strongly limit the population fraction of dark matter with such interactions, thus providing a unique opportunity to probe extremely dilute cosmological relics.
Other energy deposition mechanisms may apply~\cite{Graham:2018efk,NSvIR:magneticBH:2020,NSvIR:tidalfifthforce:Gresham2022}.
The production of GRBs could lead to other interesting signatures, {\em e.g.} a detectable fraction of NSs seen to be overluminous in sky surveys~\cite{NSvIR:clumps2021}.
The PT could also generate fast radio burst signatures~\cite{FRB:Fuller:2014rza}.
The deconfinement PT itself could be more interesting than discussed here.
An intriguing possibility is a second critical point at low temperatures in the QCD phase diagram, in which case it could be a crossover~\cite{Hatsuda:2006ps,review:QCDPD:FukushimaHatsuda:2010bq,review:BaymHatsuda:2017whm}.
Transitions to a ``quarkyonic'' phase, where quarks near the Fermi surface confine to form baryons, would be second-order~\cite{quarkyonic:McLerran:2007qj,*quarkyonic:Fukushima:2015bda, *quarkyonic:McLerranReddy:2018hbz}.
Sequential PTs to two phases of superconducting quark matter are possible~\cite{Alford:2017qgh}.
We leave to future work studies of the unique fingerprints of these scenarios.

Were the potential barrier $U_{\rm max}$ smaller than our choice of 10 GeV, interesting prospects could open up. 
Our limits can be extended down to smaller $\mx$ than displayed here, and elastic scatters may come into play, though one must be careful to account for the likelihood that the deposited energy is contained within an $\Oc$(fm) region after accounting for Pauli-blocked nucleons.
Other fundamental physics may be probed, {\em e.g.} nucleon decay for $U_{\rm max} \lsim$~GeV. 
Assuming conservatively $10^{55}$ protons and $10^{56}$ neutrons within $R_{\rm dcf}$ and Gyr-old NSs, we get lower limits of $10^{64}$~yr for the proton lifetime (c.f. $\Oc(10^{34})$~yr limits at laboratory searches~\cite{SKNucDK:2016eqm}) and $10^{65}$~yr for the inverse rate of an exotic branch of neutron decay.
Even if the decay final states do not inject energy and escape the NS, the nucleon Auger effect~\cite{NSheat:DarkBary:McKeen:2020oyr,*NSheat:Mirror:McKeen:2021jbh} from the removal of nucleons from their Fermi seas would apply for $U_{\rm max} \lsim \Oc(100)$~MeV. 
We remark that impressive limits would be obtained even if 
(a)  tunneling through a higher barrier delayed the decay,
(b) only small volumes in the NS core (where $U_{\rm max}$ lowers) are considered. 

In light of the powerful statements that can be made in fundamental physics under the right assumptions, gains in certainty on QCD/nuclear issues, such as EoS models, become crucial.

%%%%%%%
\section*{Acknowledgments}
%%%%%%%
Many thanks to 
Joe Bramante 
for remarks on the manuscript, and
Chris Dessert,
Chethan Krishnan,
Prasad Hegde,
Ranjan Laha,
Shirley Li,
Maneesha Pradeep,
Sanjay Reddy,
Akash Kumar Saha, 
Flip Tanedo,
Fridolin Weber,
and
Hai-Bo Yu
for discussions.
N.~R.~acknowledges support from the grant ANRF/ECRG/2024/000387/PMS and the Infosys Foundation, Bangalore.
Z.~Z.~is supported by the Prime Minister's Research Fellows (PMRF) scheme, Ref. No. TF/PMRF-22-7307.

\bibliography{refs}

\end{document}